\documentclass[final]{ias2}

\usepackage{graphicx} 
\usepackage{multirow}
\usepackage{array} 

\usepackage{hyperref} 

\begin{document}

\markboth{Pramana class file for \LaTeX 2e}{R. Arun, et. al.}

\title{Impact of Current on Static and Kinetic Depinning Fields of Domain Wall in Ferromagnetic Nanostrip}

\author[bdu]{R. Arun} 
\email{arunbdu@gmail.com}
\author[sastra]{P. Sabareesan} 
\email{sabaribard@gmail.com}
\author[bdu]{M. Daniel}
\email{danielcnld@gmail.com}
\address[bdu]{Centre for Nonlinear Dynamics, School of Physics, Bharathidasan University, Tiruchirapalli - 620 024, Tamilnadu, India.}
\address[sastra]{Centre for Nonlinear Science and Engineering, School of Electrical and Electronics Engineering, SASTRA University, Thanjavur - 613 401, Tamilnadu, India.}

\begin{abstract}
The impact of current on static and kinetic depinning fields of a domain wall in an one dimensional ferromagnetic nanostrip is investigated by solving the Landau-Lifshitz-Gilbert equation with adiabatic and non-adiabatic spin-transfer torques analytically and numerically. The results show that in the absence of current, the static depinning field is greater than the kinetic depinning field and both the depinning fields decrease by the increase of current applied in a direction opposite to the direction of the applied field. Both the depinning fields can also be tuned by the current to make them equal.
\end{abstract}

\keywords{Domain wall, pinning, depinning, spin transfer torque, spin polarized current, Landau-Lifshitz-Gilbert equation}

\pacs{75.60 Ch, 75.70 Kw, 75.78 Fg, 72.25 Pn}
 
\maketitle


\section{Introduction}
In the recent years, research in the motion of domain wall in ferromagnetic nanostrip has been focused more due to its practical applications such as magnetic logic\cite{Allwood} and memory devices\cite{Parkin}.  The manipulation of domain wall in magnetic nanostrip can be achieved by the application of magnetic field\cite{Schryer} and/or current\cite{Li,Zhang}. By applying the magnetic field and current, the domain wall moves rigidly up to the critical value known as Walker limit.  Above the Walker limit, the motion of the domain wall is not regular and exhibits oscillatory behavior\cite{014413,024217,093913}. Conventionally, the position of domain wall can be controlled by ion irradiation\cite{499,232402,163901}, nearby nano-magnets\cite{087204,144418,17D503,17D506} and introducing artificial geometrical constraints such as notches in the long edge of the wire\cite{102509, 033904, 262501, 054414, 024426,054423}. The displacement of a domain wall from one notch to another notch can be achieved by depinning the trapped wall from one notch and pinning the moving wall in the other notch and its pinning mechanism could be understood by static and kinetic pinning respectively.  The trapped domain wall can be depinned from a notch when the applied magnetic field is above the threshold magnetic field, which is called the static depinning field.  Similarly, the moving domain wall cannot be trapped, when it crosses the notch if the applied magnetic field is above the threshold magnetic field, which is known as the kinetic depinning field. The static depinning field and kinetic depinning field are independent of the shape of the notch whether it is rectangle or triangle in shape and the kinetic depinning field depends on the chirality of the domain wall\cite{033904,054414}. 

Instead of producing a notch by single triangle, one can create the notch by two symmetrical triangles in the nanostrip which has more advantage than the single one because of the irrelevance of static depinning field on the chirality and propagation direction of a domain wall\cite{054414}. Recently, Sung-Min Ahn $et~al$\cite{07D309,152506} have found the following results for the notch created by two symmetrical triangles. The static and kinetic depinning fields decrease with the increase of the width of the nanowire\cite{07D309} and the static depinning field is greater than the kinetic depinning field\cite{152506}. For the above results, they have modelled the pinning field created by the notch as a step function. Though there are plenty of studies in field driven domain wall pinning, the study of domain wall pinning in the presence of current is limited. When the applied current is above the threshold, the trapped domain wall in a notch can be untrapped, and the corresponding current is known as static depinning current. It increases when the notch depth\cite{113913} and notch angle\cite{127203} are increased. And also, the static depinning field reduces when the current is applied in a direction opposite to the direction of the domain wall propagation\cite{7266}.  The static depinning current increases or decreases with the increase of doping concentration or magnetic field respectively\cite{020413}.

In this paper, the dynamical equation of domain wall with a symmetrical notch has been solved analytically and numerically. The results show the variation of static and kinetic depinning fields and the equivalence of both depinning fields in the presence of current. The paper is organised as follows: The model of the domain wall along with the pinning field is discussed in Section 2 and the corresponding equations for the velocity, width and excitation angle of the domain wall are derived analytically in Section 3 in the presence of current along with the pinning field. The numerical results are explained in Section 4 and finally the results are summarized
 in Section 5.

\section{Model}
\begin{figure}[!hbtp]
\centering\includegraphics[angle=0,width=0.6\linewidth]{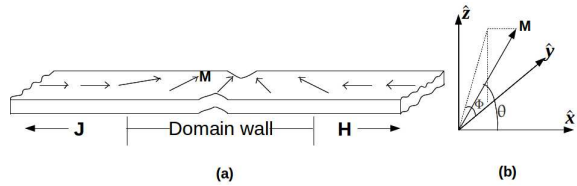}
\caption{ (a) A Schematic diagram of the ferromagnetic nanostrip having an artificial notch in the form of two symmetrical triangles which is taken as our model and its magnetization varies only along x-direction.  The current density ${\bf J}$ and external magnetic field ${\bf H}$ have been applied in negative x-direction and positive x-direction respectively. ~(b) Representation of magnetization in terms of spherical coordinates.}
\end{figure}

We consider an infinitly long ferromagnetic nanostrip with a single Neel-type transverse domain wall as shown in FIG.1(a) as our model to study the phenomena of static and kinetic pinning of the domain wall. The pinning field is introduced by etching the center of the nanostrip with an artificial notch which is in the form of symmetrical triangles as shown in FIG.1(a). This symmetrical nature of the notch is more advantageous than the asymmetrical notch because of its ability to create an unique static depinning field irrespective of the domain wall polarity and its propagation direction. The two domains of the ferromagnetic nanostrip are separated by a domain wall with opposite magnetizations(${\bf M}$) along the easy axis(x-axis) and the arrows in the strip show the direction of magnetization.  The external time varying magnetic field(${\bf H}(t)$) and current(${\bf J}(t)$) are applied along the positive and negative x-direction respectively.  In FIG.1(b), ${\bf \hat x,\hat y,\hat z}$ represent the unit vectors along x,y and z directions respectively and the angles $\theta$ and $\Phi$ refer to the angle between the magnetization vector ${\bf M}$  and the positive x-direction and the angle between the projection of ${\bf M}$ in yz-plane and positive y-direction respectively. Assume that the variation of ${\bf M}$ takes place only along the x-direction and the angles $\theta$ and $\Phi$ imply the deviation of the magnetization vector from positive x-direction and xy-plane respectively. The dynamics of domain wall is understood through the spatial and time variation of magnetization governed by the Landau-Lifshitz-Gilbert (LLG) equation in the presence of field and current.  The LLG equation is written as\cite{Zhang}
\begin{subequations}
\label{LLGeq}
\begin{align}
&\frac {\partial {\bf M}(x,t)}{\partial t}  =  -\gamma {\bf M} \times {\bf H}_{eff} + \frac{\alpha }{M_{s}} {\bf M} \times \frac{\partial {\bf M}}{\partial t} -\frac{b(t)}{M_{s}^2} {\bf M} \times \left({\bf M} \times \frac{\partial {\bf M}}{\partial x}\right)\nonumber\\&~~~~~~~~~~~~~~~~~~~~~~- \frac {c(t)}{M_{s}} {\bf M} \times  \frac{\partial {\bf M}}{\partial x}, \label{LLG}\\
&{\bf M}=(M_x,M_y,M_z);~|{\bf M}|^2=M_x^2+M_y^2+M_z^2=M_s^2, \label{LLG1}
\end{align}
\end{subequations}
where,
\begin{subequations}
\label{stt}
\begin{align}
&b(t)=\frac{P \mu_B J(t)}{ e M_s}, \label{b}\\
&c(t)=\xi b(t).
\end{align}
\end{subequations}
Here, $\gamma$ is the gyromagnetic ratio, $\alpha$ is the Gilbert damping parameter, $M_s$ is the saturated magnetization, ${\bf H}_{eff}$ is the effective field.  $b(t)$ and $c(t)$ represent the magnitude of adiabatic and non-adiabatic spin-transfer torques respectively, which include the interaction of conduction electrons and the local magnetization.  $P$ is the polarization, $J$ is the magnitude of current density, $\mu_B$ is the Bohr magneton, $e$ is the charge of electron and $\xi(\approx 0.01)$\cite{Zhang} is the ratio between $c$ and $b$.    The adiabatic spin-transfer torque is corresponding to the reaction torque on the magnetization produced by the spatial variation of the spin current density\cite{Zhang1}.  And the non-adiabatic spin-transfer torque corresponds to the reaction torque on the magnetization due to the continuous space variation of spatially mistraking spins between conduction electrons and local magnetization\cite{Zhang}.  The initial and final velocity of a domain wall is controlled by adiabatic and non-adiabatic spin-transfer torques respectively\cite{Li,Zhang}.
  The effective field ${\bf H}_{eff}$ includes the fields due to exchange energy, easy axis anisotropy, external field, demagnetization field and the pinning field produced by the notch\cite{152506} is given by
\begin{align}
{\bf H}_{eff} &= \frac{2A}{M_{s}^2} ~\frac{\partial^2 {\bf M}}{\partial x^2} + \left(\frac{H_{k}}{M_{s}}M_{x}+H(t)-H_p U(x)\right) {\bf\hat x}  - 4\pi M_{z} {\bf \hat z}, \label{Heff}
\end{align}
where,
\begin{subequations}
\begin{align}
U(x) &= 1,~\rm{for} ~0<x\leq\delta, \label{U}\\
    &= 0,~\rm{otherwise}. \label{U1}
\end{align}
\end{subequations}
Here, $A$ represents the exchange interaction coefficient, $H_k$ represents the magnetocrystalline anisotropy coefficient, $H$ is magnitude of external magnetic field and $4\pi M_z$ is the demagnetization field.  $H_p$ is the pinning field, $U(x)$ is the step function used to introduce the pinning field in the region $0<x\leq\delta$ and $\delta$ is the range of the pinning field.  

\section{Analytical solutions for domain wall parameters}

In order to understand the dynamics of the domain wall in pinning field, we try to solve Eq.\eqref{LLG}.  As Eq.\eqref{LLG} is a highly nontrivial vector nonlinear evolution equation, it may be difficult to solve the same in its present form.  Hence, we rewrite Eq.\eqref{LLG} in terms of the polar coordinates by using the transformations $M_{x} = M_{s} \cos\theta,M_{y} = M_{s} \sin\theta\cos\Phi,M_{z} = M_{s} \sin\theta\sin\Phi$ and the resultant equations read

\begin{subequations}
\label{llg}
\begin{align}
\frac{\partial \theta}{\partial t} + \alpha \sin\theta \frac{\partial \Phi}{\partial t} &=~  \frac{2\gamma A}{M_s}\left(2 \cos\theta \frac{\partial \theta}{\partial x} \frac{\partial \Phi}{\partial x} + \sin\theta \frac{\partial^2 \Phi}{\partial x^2} \right)\notag\\ - &2\pi{\gamma} M_s \sin\theta \sin2\Phi + b(t) \frac{\partial \theta}{\partial x} + c(t) \sin\theta \frac{\partial \Phi}{\partial x}, \label{llg1} \\
\alpha\frac{\partial \theta}{\partial t}- \sin\theta \frac{\partial \Phi}{\partial t} &=~ \frac{2\gamma A}{M_{s}}\left(\frac{\partial^2 \theta}{\partial x^2}-\sin \theta \cos\theta\left(\frac{\partial \Phi}{\partial x}\right)^2\right)\notag\\
& - \gamma [H(t)-H_p U(x)] \sin\theta  -\frac{\gamma}{2}\left(H_k + 4\pi M_s \sin^2\Phi\right)\sin2\theta\nonumber\\
& - b(t) \sin\theta \frac{\partial \Phi}{\partial x} + c(t)\frac{\partial \theta}{\partial x}. \label{llg2}
\end{align}
\end{subequations}
Eqs.\eqref{llg} can be solved by using the following trial functions introduced by Schryer and Walker\cite{Schryer,Li}.
\begin{subequations}
\label{trialfunctions}
\begin{align}
\theta(x,t) &= 2 \tan^{-1}\exp\left(\frac{x-X(t)}{W(t)}\right), \label{trial1}\\
\Phi(x,t) &= \phi(t), \label{trial2}
\end{align}
\end{subequations}
where, $X(t)$ is the position of the center of the domain wall and $W(t)$ is the width of the domain wall.  Eq.\eqref{trial1} assumes that as time goes on, the domain wall moves without any change in the static profile(the spatial variation of $\theta$ in the absence of current and field) and only with the change in its width.  And Eq.\eqref{trial2} assumes that the domain wall excites from the xy-plane with time and the excitation is independent of space.  $\phi(t)$ is an angle between the projection of magnetization(yz-plane) and positive y-direction, which can be called as excitation angle. 

On substituting $\frac{\partial\theta}{\partial t}$ from Eq.\eqref{llg2} in Eq.\eqref{llg1}, we get
\begin{align}
&{(1+\alpha^2)\sin\theta }\frac{\partial \Phi}{\partial t} = ~\frac{2A\gamma}{M_s}\left[\alpha\sin\theta~\frac{\partial^2 \Phi}{\partial x^2} + 2\alpha\cos\theta~\frac{\partial \theta}{\partial x}\frac{\partial \Phi}{\partial x}-\frac{\partial^2 \theta}{\partial x^2}+\sin\theta~\cos\theta~\left(\frac{\partial \Phi}{\partial x}\right)^2\right] \nonumber\\
&~~~~~~~~~-2\alpha\gamma\pi M_s\sin2\Phi~\sin\theta + \frac{\gamma}{2}(H_k+4\pi M_s\sin^2\Phi)\sin2\theta+ \gamma [H(t)-H_p U(x)] \sin\theta  \nonumber\\ 
&~~~~~~~~~ +(1+\alpha\xi)b(t)~\sin\theta ~\frac{\partial \Phi}{\partial x} + (\alpha-\xi)b(t)~\frac{\partial \theta}{\partial x}.
\label{eq1}
\end{align}
Similarly, by substituting $\frac{\partial\Phi}{\partial t}$ from Eq.\eqref{llg1} in Eq.\eqref{llg2}, we get
\begin{align}
&{(1+\alpha^2)}\frac{\partial \theta}{\partial t} =\frac{2A\gamma}{M_s}\left[\sin\theta ~\frac{\partial^2 \Phi}{\partial x^2}+ 2\cos\theta~ \frac{\partial \theta}{\partial x}\frac{\partial \Phi}{\partial x} +\alpha\frac{\partial^2 \theta}{\partial x^2}-\alpha\sin\theta~\cos\theta~\left(\frac{\partial \Phi}{\partial x}\right)^2\right]\nonumber\\ 
&~~~~~~~~~- 2\gamma\pi M_s\sin2\Phi~\sin\theta- \frac{\gamma\alpha}{2}\left[H_k+4\pi M_s\sin^2\Phi\right]\sin2\theta -\alpha\gamma [H(t)-H_p U(x)]\sin\theta\nonumber\\
&~~~~~~~~~  -(\alpha-\xi)b(t)~\sin\theta~\frac{\partial \Phi}{\partial x} + (1+\alpha\xi)b(t)~\frac{\partial \theta}{\partial x}. \label{eq2}
\end{align}

From Eqs.\eqref{trial1} and \eqref{trial2}, one can derive the following identities at $x = X$.
\begin{subequations}
\label{identities}
\begin{align}
&\theta(X,t) = \pi\label{identity1},\\
&\frac{\partial \theta (X,t)}{\partial x} = \frac{1}{W(t)}, \label{identity2} \\
&\frac{\partial^2 \theta (X,t)}{\partial x^2} = 0, \label{identity3}\\ 
&\frac{\partial^3 \theta (X,t)}{\partial x^3} = - \frac{1}{{W(t)^3}} ,  \label{identity4} \\
&\frac{\partial \theta (X,t)}{\partial t}=-~\frac{1}{W(t)}\left(\frac{dX}{dt}\right), \label{identity5}\\
&\Phi(X,t) = \phi(t),\label{identity6}\\
&\frac{\partial \Phi (X,t)}{\partial t} = \frac{d\phi(t)}{dt},\label{identity7}\\
&\frac{\partial \Phi (X,t)}{\partial x} = \frac{\partial^2 \Phi (X,t)}{\partial x^2} =  \frac{\partial^2 \Phi (X,t)}{\partial x \partial t} = 0.\label{identity8} 
\end{align}
\end{subequations}
The reduced form of Eq.\eqref{eq1} is obtained by substituting $x=X(t)$ in it.  Further, it can be simplified by using the identities given in Eqs.\eqref{identities} and we can obtain,
\begin{align}
(1+\alpha^2)\frac{d\phi(t)}{dt} = &\gamma (H(t)-H_p U(X)-2\pi\alpha M_s\sin2\phi ) +\frac{(\alpha-\xi)b(t)}{W(t)} . \label{dphi}
\end{align}
Similarly from Eq.\eqref{eq2} we can derive
\begin{align}
v(t)=\frac{dX}{dt} = \frac{\gamma W(t)}{(1+\alpha^2)}{\left(2\pi M_s\sin2\phi + \alpha [H(t)-H_p U(X)]\right)}- b(t)\left(\frac{1+\alpha\xi}{1+\alpha^2}\right)~. \label{v}
\end{align}
The width of the domain wall is obtained by differentiating Eq.\eqref{eq1} with respect to $x$ and reducing it at $x=X(t)$ and the reduced equation can be simplified using the identities given in Eqs.\eqref{identities}.
\begin{align}
W(t) ={W_0}\left[1 + ({4\pi M_s}/{H_k}) \sin^2\phi  \right]^{-\frac{1}{2}}. \label{W}
\end{align}
Here, $W_0=\sqrt{2A/H_k M_s}$ is the initial width of the domain wall. Eqs.\eqref{dphi}, \eqref{v} and \eqref{W} give the variation of the excitation angle, velocity and width with respect to time respectively in the presence of current and external field as time passes.  Since Eqs.\eqref{v} and \eqref{W} involve $\phi$, in order to find the velocity and width of the domain wall, it is needed to solve Eq.\eqref{dphi}.
  As Eq.\eqref{dphi} is a highly nontrivial nonlinear evolution equation it is difficult to solve the same analytically.  Hence, we solve Eq.\eqref{dphi} numerically and the results are discussed in the forthcoming sections. 

\section{Numerical Results}
The pinning and depinning of the domain wall can be understood numerically by integrating the dynamical equation \eqref{dphi} for the excitation angle $\phi$ by using Runge-Kutta-4 algorithm with the initial conditions $\phi(0)=0$ and using the experimentally measured values of the material parameters for Cobalt nanostripes as given by $M_s=14.46\times10^5~Am^{-1},~M_s={1.8\times10^4}/{4\pi}~Oe,~A=2\times10^{-11}~Jm^{-1},~\gamma=1.9\times10^7~Oe^{-1}s^{-1},~H_k = 500~Oe$ and $P=0.35$\cite{Li}.  The time varying field and current density are taken in the following form.
\begin{subequations}
\label{timevarying}
\begin{eqnarray}
H(t) &= (H'/h)t~&\rm{for}~0\leq t\leq h, \label{timevarying1}\\
     &=  H'~&\rm{when}~t>h, \label{timevarying2}\\
J(t) &= (J'/h)t~&\rm{for}~0\leq t\leq h, \label{timevarying3}\\
     &=  J'~&\rm{when}~t>h, \label{timevarying4}\\
h    &=  10^{-12}.& \nonumber
\end{eqnarray}
\end{subequations}
Eqs.\eqref{timevarying} represent the external field $H$ and the current density $J$ in the form of linearly increasing pulse with the duration of 1 ps and after that both are maintained as constant.  $H'$ and $J'$ can be referred as the saturated external field and current density respectively.    If the value of $H'$ or $J'$ is just above the static depinning value, the wall is depinned from the pinning field region and travels along the direction of the field (or) along the direction opposite to the direction of the current. To avoid the non-zero initial velocity of the domain wall when it enters the pinning field region to find the static depinning field, the values of $H$ and $J$ are increased from zero.  Otherwise, there will be no difference between the kinetic and static depinning. $h$ decides the rate of change of the field and current which has been fixed as 1 ps throughout this paper. In order to understand the static and the kinetic depinning of field and current, the displacement $X(t)$ of the domain wall is obtained numerically by applying Simpsons's 3/8 rule as follows.
\begin{align}
X(t) = \int_0^t v(t) ~dt. \label{X}
\end{align}

\subsection{Static pinning and depinning under field and current:}
In this section, the details of the systematic investigation of the static pinning and depinning of a domain wall in the presence of either field or current are discussed. Initially, the domain wall is placed at a position $x=0$, just before the pinning field region ($0<x\leq\delta$) with a pinning field of strength $H_p$=100 Oe. The domain wall is moved from rest by applying a time varying field $H$(saturated field: $H'$) along the positive x-direction or by applying a current density $J$(saturated current density: $J'$) along the negative x-direction.
To understand whether the wall moves beyond the pinning field region or not under field(current), the displacement of the domain wall is numerically calculated from Eq.\eqref{X} for $H'(J')$.  If the strength of the saturated field $H'$ (or) current density $J'$ is not sufficient to move the domain wall beyond the pinning field region, then the wall is trapped or pinned within the pinning field region created by the notch in the nanostrip.  

	The displacement of the wall $X(t)$ is obtained numerically from Eq.\eqref{X} for different strengths of  saturated external field namely $H'$= 99.9 Oe, 100.0 Oe and 100.1 Oe for a pinning field strength $H_p$=100.0 Oe and for a pinning field range of $\delta$=$W_0$=23.52 nm is shown in FIG.2(a).  When $H'<=$ 100.0 Oe,  the domain wall is pinned in the pinning field region and the corresponding displacement is constant.  Once the field $H'$ is just above the 100.0 Oe, the wall crosses the pinning field region and displaces linearly as time passes on as shown in FIG.2(a), which indicates that the static depinning field($H_{sdp}$) of the wall driven by field in the absence of current is 100.0 Oe and it is equal to the given pinning field strength. In a similar way, the static depinning current density($J_{sdp}$) is understood from FIG.2(b), where the plots showing the displacement of the domain wall against time have been plotted for different values of saturated current density: $J'$ = -58.7$\times$10$^{8}$ A/cm$^{2}$, -58.8$\times$10$^{8}$ A/cm$^{2}$ and -58.9$\times$10$^{8}$ A/cm$^{2}$ for $H_p$=100 Oe and $\delta$= 23.52 nm.  For the saturated current densities -58.7$\times$10$^{8}$ A/cm$^{2}$ and -58.8$\times$10$^{8}$ A/cm$^{2}$, the wall is pinned in the pinning field region, whereas for $J'$=-58.9$\times$10$^{8}$ A/cm$^{2}$ the wall moves out of the pinning field region as shown in FIG.2(b).  Hence, the static depinning current density $J_{sdp}$ of the wall driven by current when the field is absent is given by -58.8$\times$10$^{8}$ A/cm$^{2}$.

\begin{figure}[!h]
\centering\includegraphics[angle=0,width=0.5\linewidth]{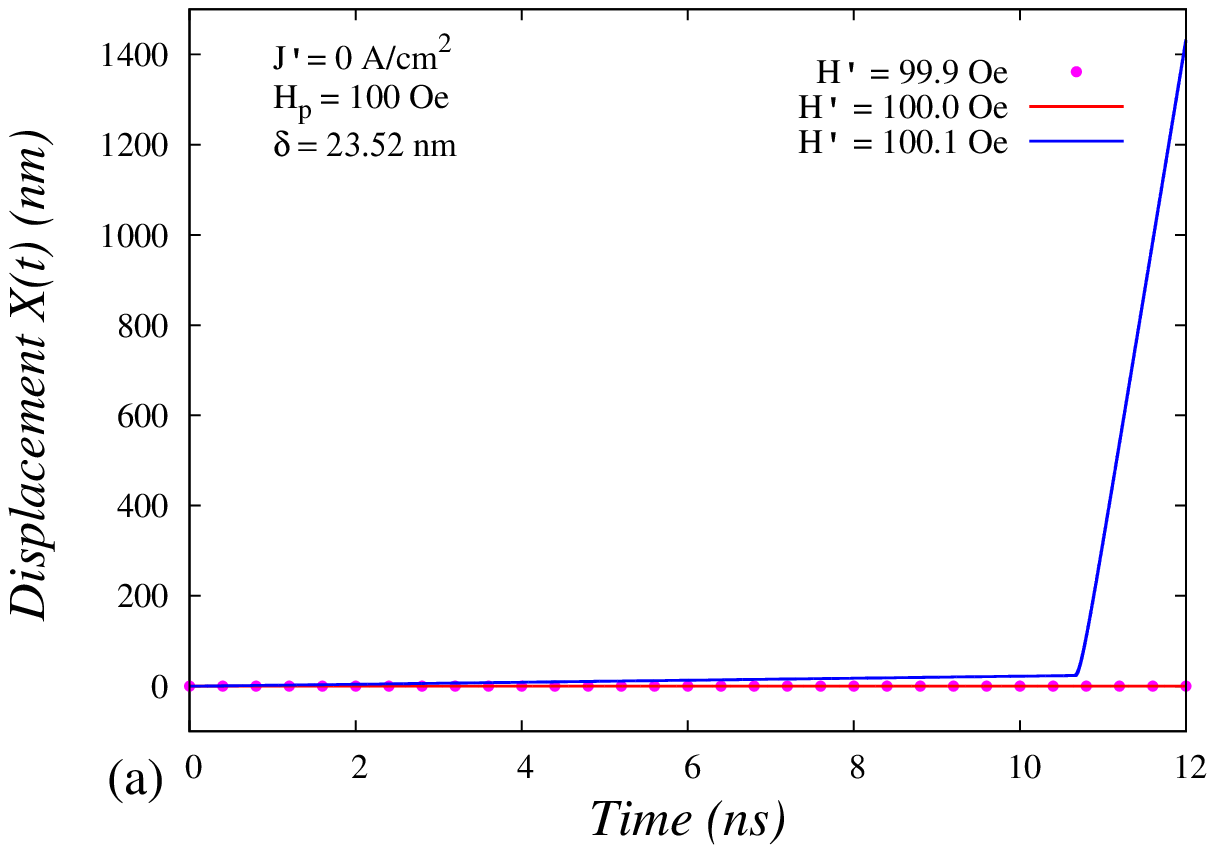}~\includegraphics[angle=0,width=0.5\linewidth]{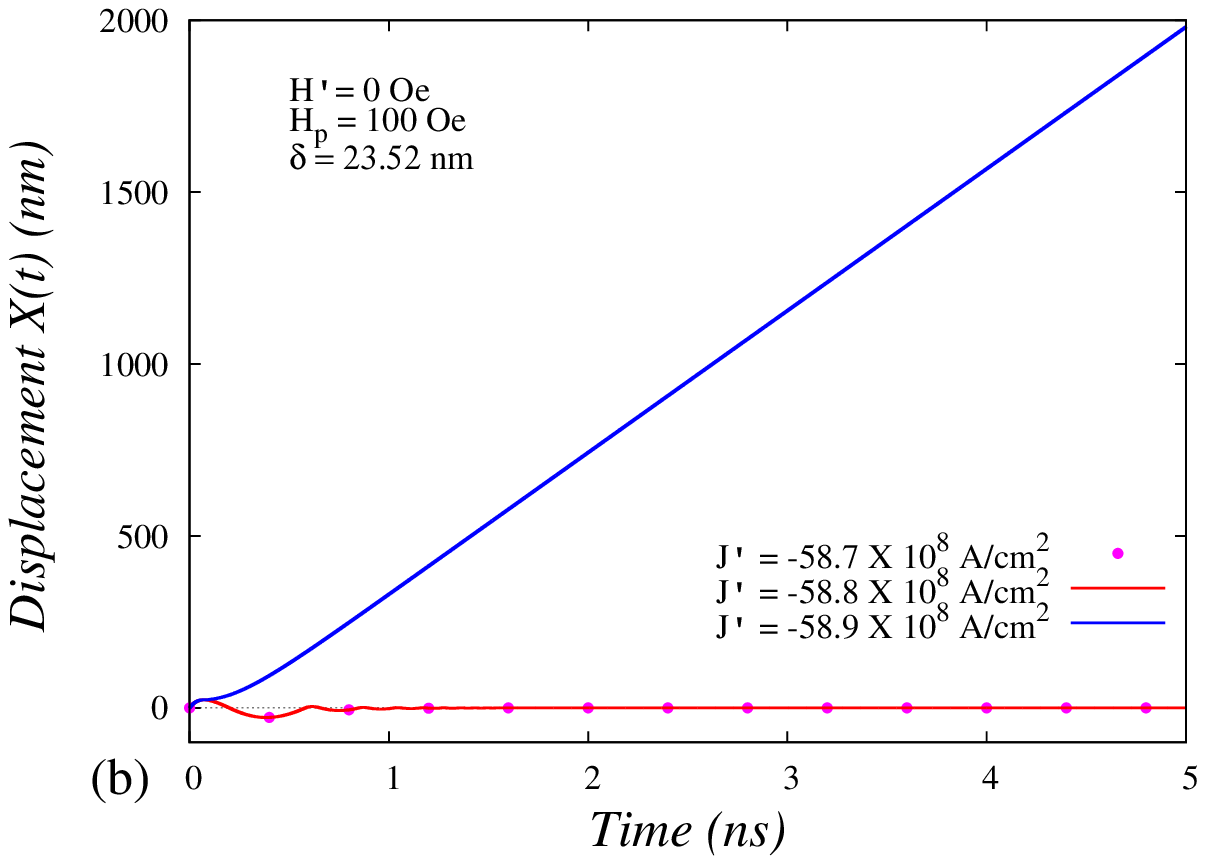}
\caption{(color online).  Displacement $X(t)$ of the domain wall against time for (a) field driven and (b) current driven cases. The pinning field strength is fixed as $H_p$=100 Oe and the range of the pinning field as $\delta=W_0$=23.52 nm.}
\end{figure}

\subsection{Kinetic pinning and depinning under field and current:}
In the case of static pinning and depinning, the domain wall is moved from rest into the pinning field region whereas the kinetic pinning and depinning of the field and current are studied by allowing the moving domain wall into the pinning field region. Therefore, the initial position of the wall is taken far away from the pinning field region along negative x-direction in such a way that the domain wall moves with constant velocity when it enters into the pinning field region.  The strength of pinning field and its range is kept as same as the static case. By applying the field(current) with their saturated values of $H'$($J'$), the domain wall starts to move and reaches the pinning field region with constant velocity.    The wall would not be pinned if $H'$($J'$) is just above a threshold value, otherwise it would be pinned. This threshold value of field and current are referred to kinetic depinning field($H_{kdp}$) and kinetic depinning current density($J_{kdp}$) respectively.  

\begin{figure}[!h]
\centering\includegraphics[angle=0,width=0.5\linewidth]{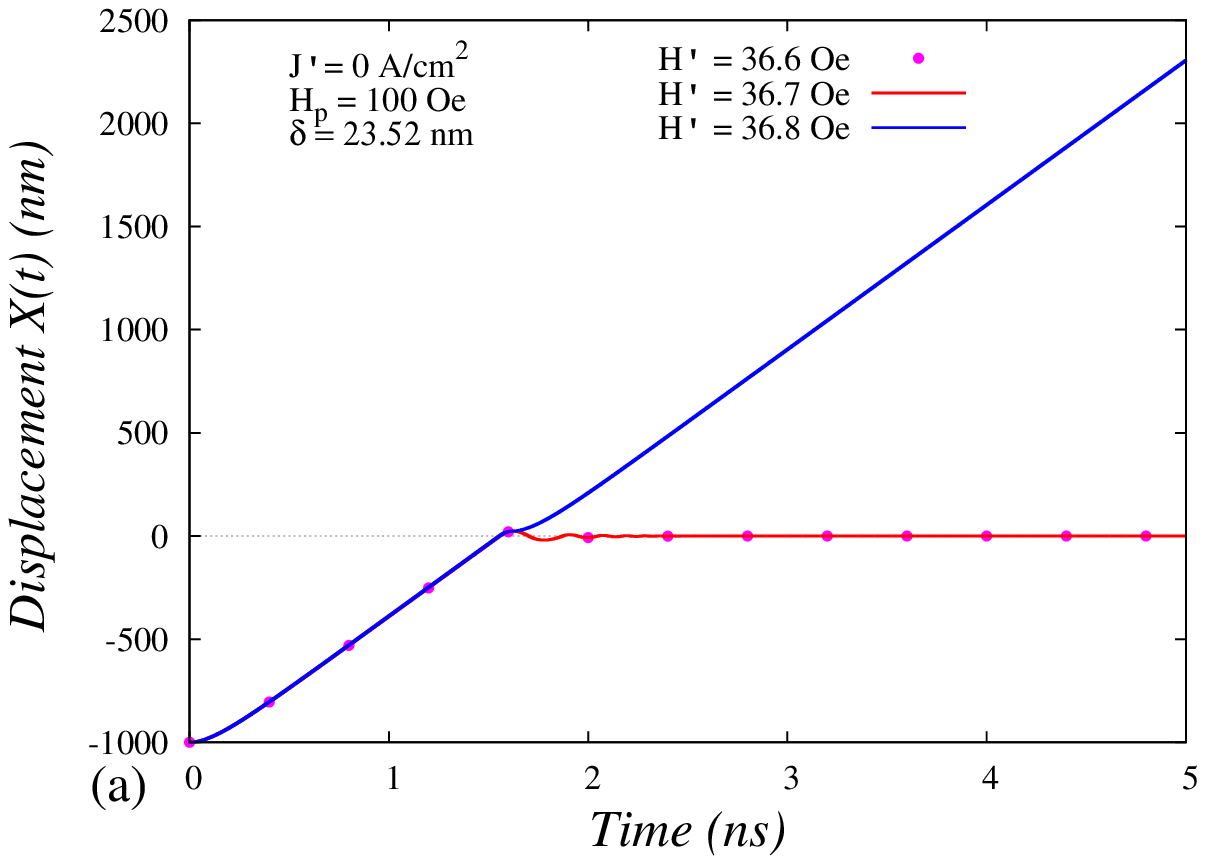}~\includegraphics[angle=0,width=0.5\linewidth]{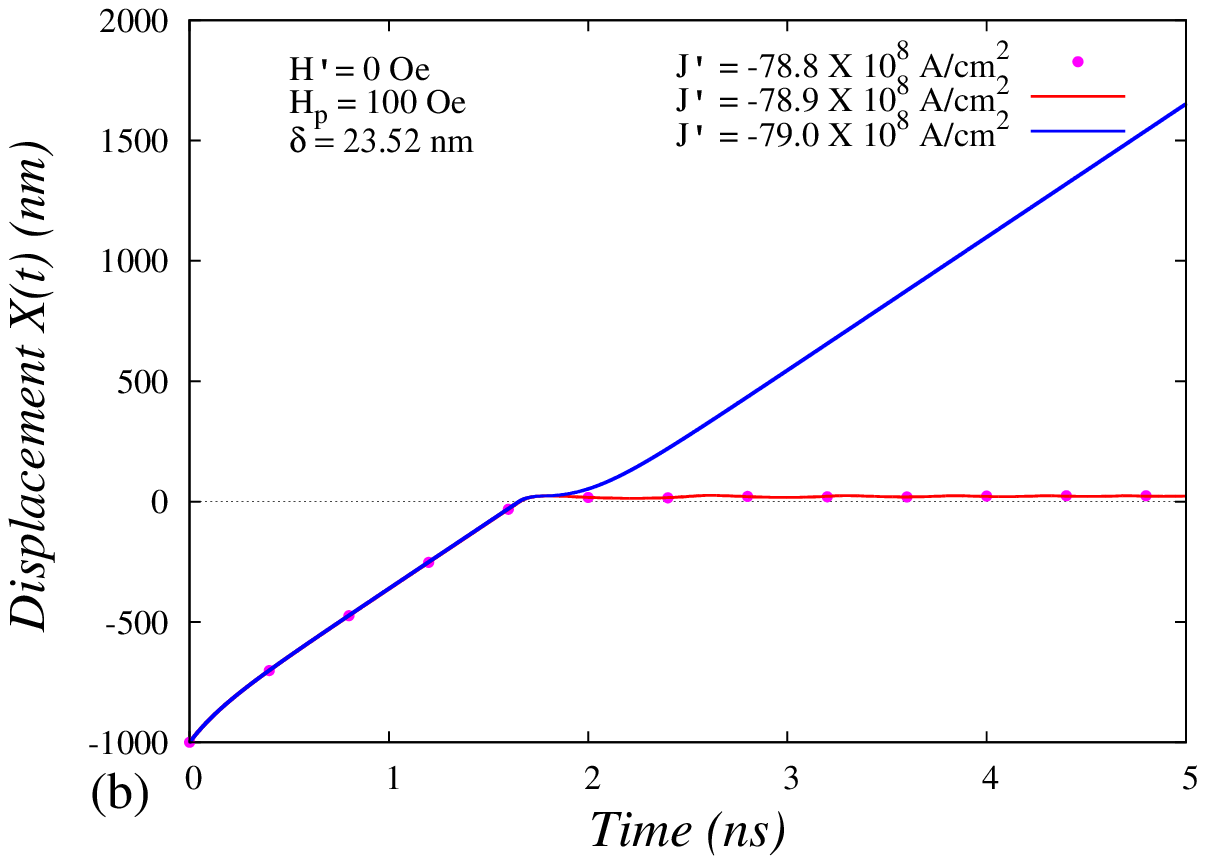}\\
\centering\includegraphics[angle=0,width=0.5\linewidth]{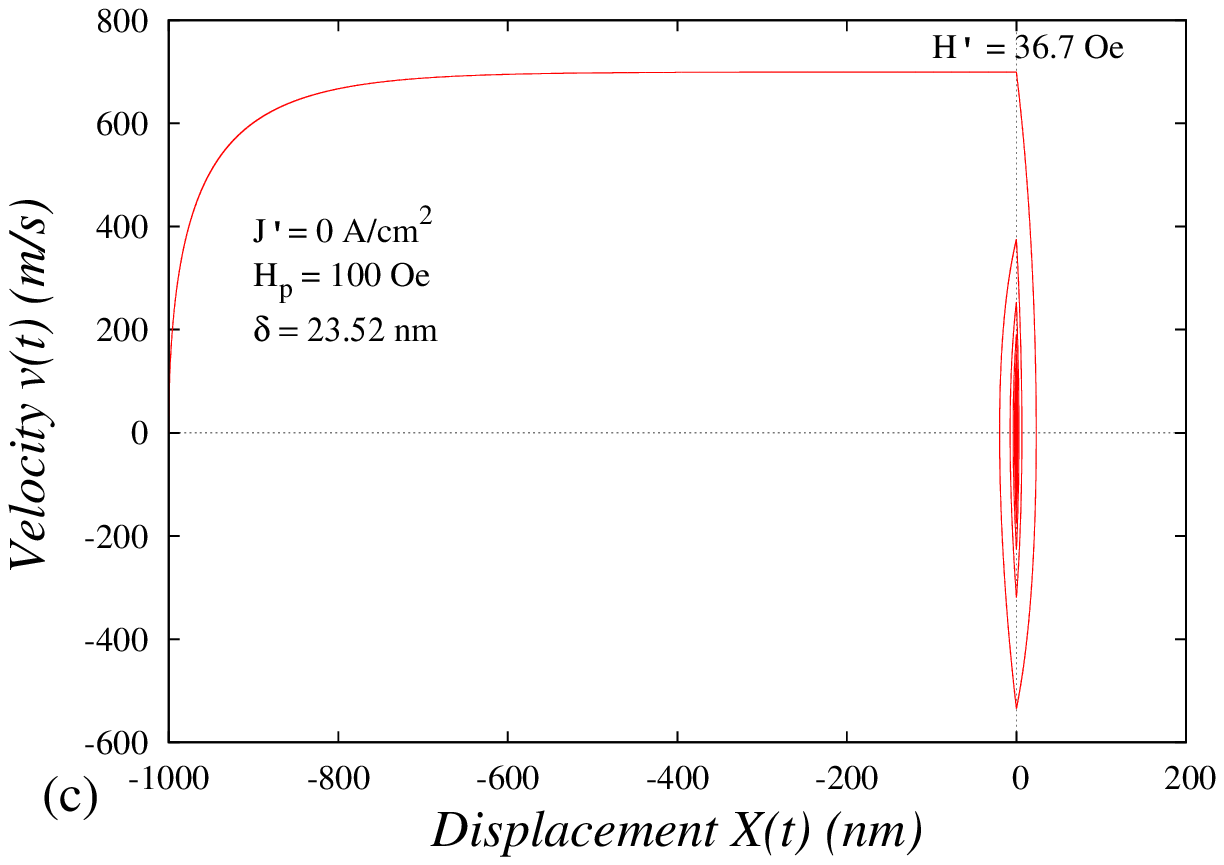}~\includegraphics[angle=0,width=0.5\linewidth]{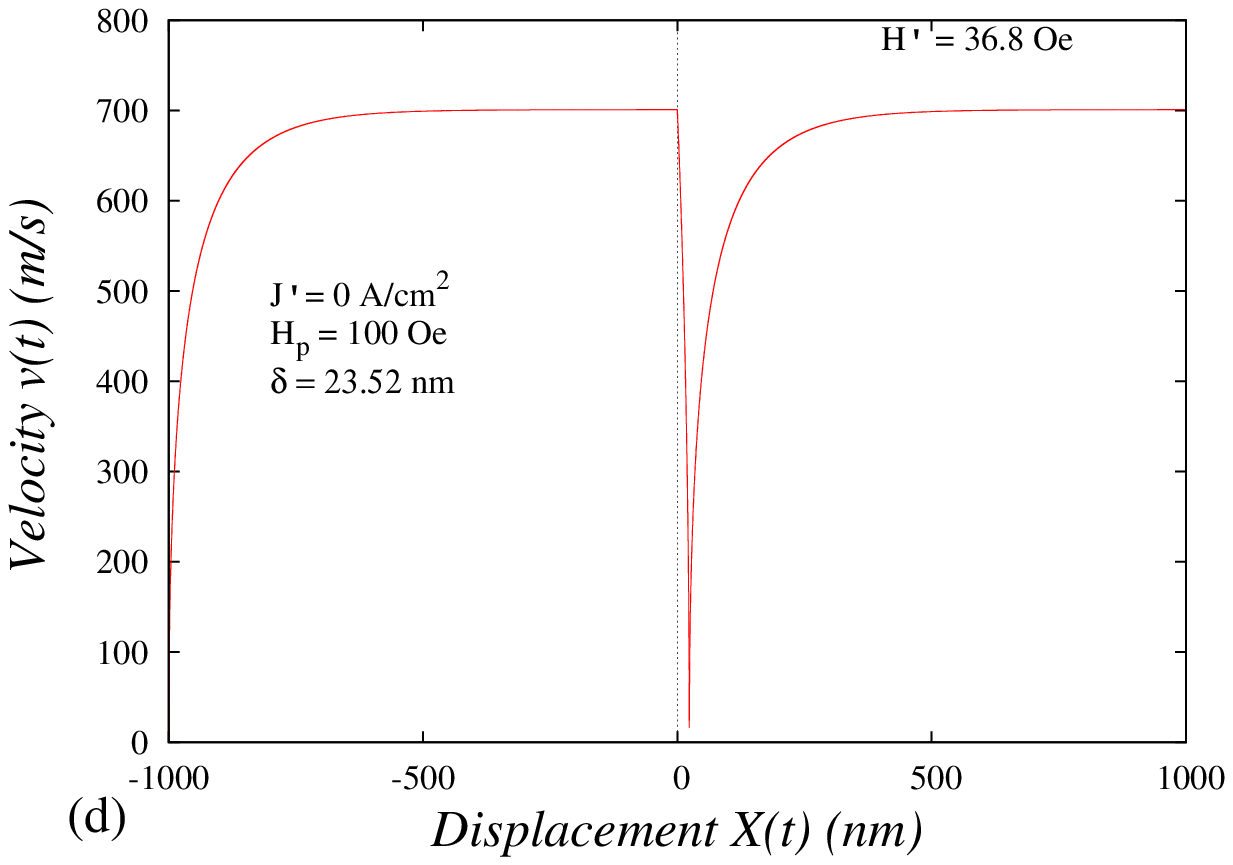}\\
\centering\includegraphics[angle=0,width=0.5\linewidth]{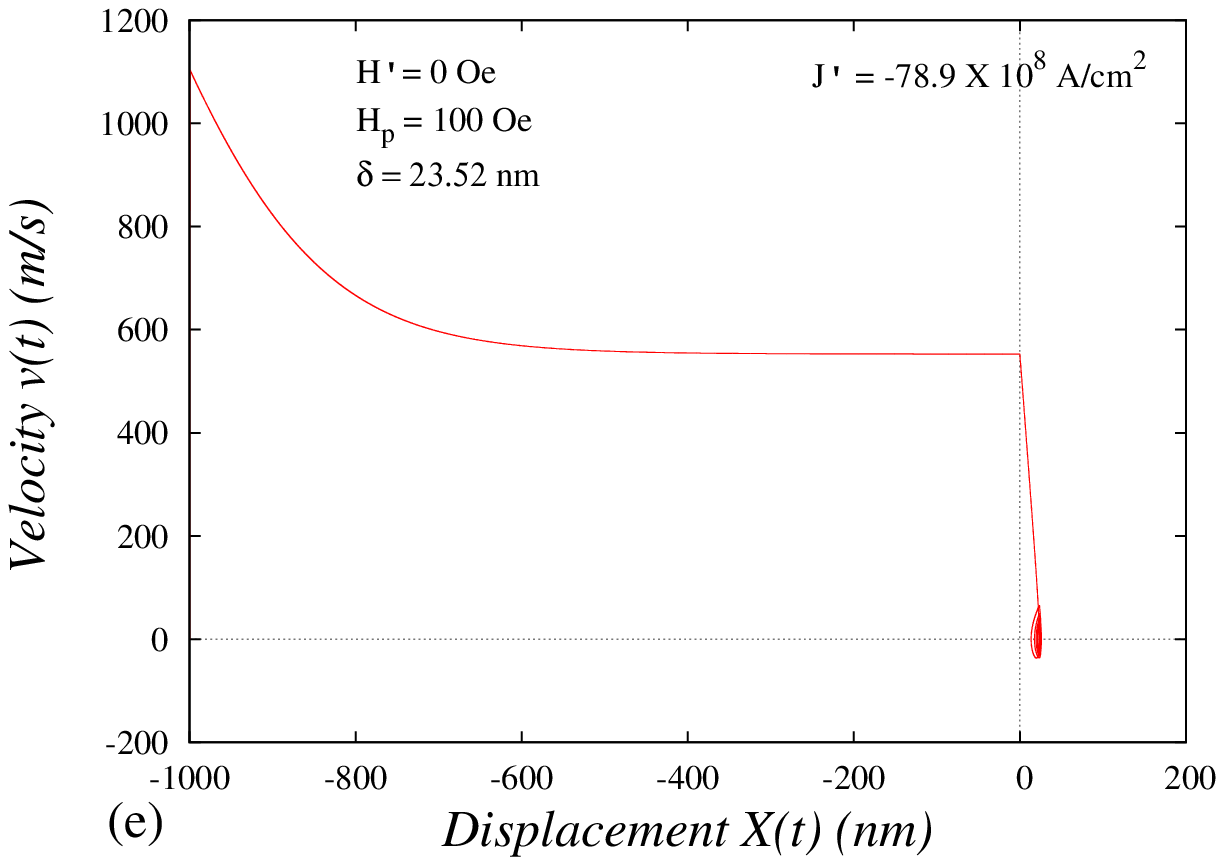}~\includegraphics[angle=0,width=0.5\linewidth]{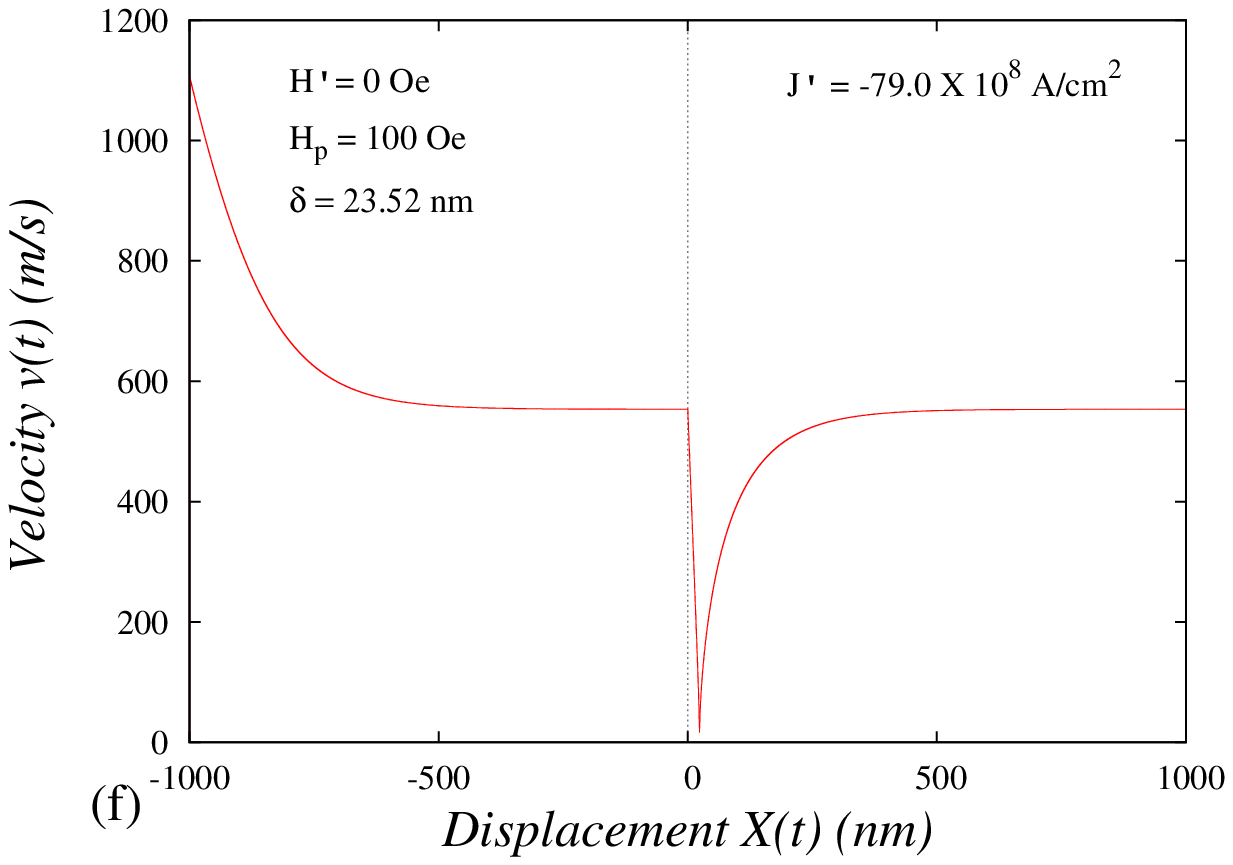}
\caption{(color online).  Displacement of the wall against time for (a) field and (b) current driven cases.  The velocity of the wall with respect to the displacement (c) at the depinning field, (d) above the depinning field for the field driven case and (e) at the depinning current density, and (f) above the depinning current density for the current driven case.  The strength of the pinning field and the range are set to be $H_p=$100 Oe and $\delta=23.52$ nm for all the plots.}
\end{figure}

The displacement($X$) of the domain wall versus time has been plotted for the field driven and current driven cases, are shown in FIGs.3(a) and 3(b) respectively.  The initial position of the domain wall is set as $X(0)$=-1000 nm and the strength and range of the pinning field is considered as 100 Oe and $\delta = 23.52$ nm respectively. In the absence of current, when the saturated field is equal to or below 36.7 Oe, the wall moves from $x$=-1000 nm and settles at 0 nm where the notch(pinning field region) is located.  When the saturated field $H'$ is increased to 37.8 Oe, the domain wall moves beyond the pinning field region as shown in FIG.3(a). In a similar way, in the absence of the field, when the saturated current density is equal to or below the value of -78.9x10$^{8}$ A/cm$^2$, the displacement of the wall X(t) settles in the notch at 0 nm.  When the saturated current density $J'$ is increased to -79.0x10$^{8}$ A/cm$^2$, the  domain wall moves beyond the pinning field region as shown in FIG.3(b). From both the figures (3a and 3b), one can observe that the kinetic depinning field $H_{kdp}$ and kinetic depinning current density $J_{kdp}$ are 36.7 Oe and -78.9$\times10^{8}$ A/cm$^2$ respectively. It is interesting to observe from the static and kinetic cases, in the absence of current, the static depinning field is greater than the kinetic depinning field whereas in the absence of field the static depinning current density is lower than the kinetic depinning current density.

The pinning and depinning of the moving domain wall in the pinning field region can be observed from figures 3(c)-(f) plotted between the  velocity and the displacement of the domain wall. FIG.3(c) and 3(d) show the plots corresponding to the saturated fields $H'=H_{kdp}$ and $H'>H_{kdp}$  in the absence of current respectively.  When $H'$ is equal to the kinetic depinning field(36.7 Oe), the velocity of the wall attains the constant value before reaching the pinning field region $0<x\leq 23.52$ nm and then it drastically drops and oscillates in the pinning field region and finally, the velocity reaches zero at $x$=0 nm and the domain wall comes to rest as shown in FIG.3(c). When the field $H'$ is increased from 36.7 Oe to 36.8 Oe, the velocity of the wall drops quickly when it enters into the pinning field region and regains the same velocity when it comes out of the pinning field region, which is observed from FIG.3(d). Similarly, the velocity of the domain wall against displacement is plotted for the saturated current densities equal to and above the kinetic depinning current density($J_{kdp}$=-78.9 $\times10^{8}$ A/cm$^2$) in the absence of field as shown in FIGs.3(e) and 3(f) respectively. FIG.3(e) shows that when $J'=J_{kdp}$ the velocity of the domain wall decreases to zero when it enters into the pinning field region $0<x\leq 23.52$ nm.  When $J'$ is just above the kinetic depinning current density (-79.0 $\times10^{8}$ A/cm$^2$),  the velocity of the domain wall decreases towards zero but it does not reach zero in the pinning field region and it retains back its velocity after crossing the pinning field region which is shown in FIG.3(f). The velocity of the domain wall after crossing the pinning field region is found as 700.8 m/s for $H'$=36.8 Oe which is slightly above the kinetic depinning field whereas in the current driven case the velocity is found as 553.0 m/s for $J'$=-79.0 $\times10^{8}$ A/cm$^2$ which is slightly above the value of the kinetic depinning current density. The velocity of the domain wall to cross the notch is less in the current driven case compared to the field driven case.

\subsection{Effect of current on the static and kinetic depinning field:}
The variation of the kinetic depinning field $H_{kdp}$ and the static depinning field $H_{sdp}$ with respect to the saturated current density $J'$  is discussed here.  In order to understand the static and kinetic depinning fields numerically, the initial position of the domain wall is fixed at 0 nm and -1000 nm respectively for a fixed saturated current density.  The saturated field $H'$ is increased from 0 by an increment of 0.1 Oe.  For each and every increment of $H'$ Eq.\eqref{X} is solved numerically and the values of static and kinetic depinning fields are obtained. This can also be repeated by increasing $J'$ from 0 to -80$\times$10$^{8}$ A/cm$^{2}$ with an increment of -0.5$\times$10$^{8}$ A/cm$^{2}$ and the results have been plotted in FIG.4.  For both the cases of static and kinetic pinning, $H_p$ and the range $\delta$ are fixed as 100 Oe and 23.52 nm respectively.
\begin{figure}[!h]
\centering\includegraphics[angle=0,width=0.8\linewidth]{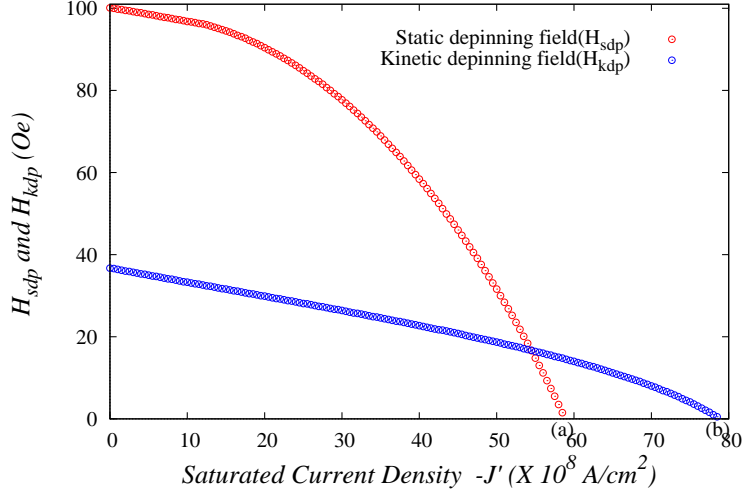}
\caption{(color online).  The static depinning field(red) and the kinetic depinning field(blue) with respect to the saturated current density($J'$) with $H_p$=100 Oe and $\delta=23.52$ nm.}
\end{figure}
    In the absence of current, the kinetic and static depinning fields are found to be 36.7 Oe and 100.0 Oe respectively.  This proves that the magnitude of the static depinning field is greater than the kinetic depinning field which has been evidenced in the earlier work\cite{152506}.  When the saturated current density is increased in the opposite direction of the external field, both the depinning fields decrease, especially the static depinning field decreases quickly than the kinetic depinning field. At a particular value of the saturated current density(=-54.47$\times$10$^{8}$ A/cm$^{2}$), both the static and kinetic depinning fields get equal value (16.7 Oe) and above this value of $J'$, the static depinning field takes the lower value than the kinetic depinning field. The decrease in static depinning field with the increase of $|J'|$ when current is applied in the opposite direction of field implies that a pinned domain wall can be depinned with the lower external field in the presence of current. Similarly, the decrease in kinetic depinning field implies that the moving wall can cross the notch with smaller external field in the presence of current. 
 From FIG.4, one can understand that the increase in saturated current density decreases both the depinning fields, and also makes the kinetic depinning field to be greater than the static depinning field.  For example, when $J'$=-52.0$\times$10$^{8}$ A/cm$^{2}$, the static depinning field is 25.2 Oe and the kinetic depinning field is 17.8 Oe, which implies that the static depinning field is greater than the kinetic depinning field.  However, when $J'$=-58.0$\times$10$^{8}$ A/cm$^{2}$, the static and kinetic depinning fields are 3.5 Oe and 15.0 Oe respectively, which implies that the static depinning field is smaller than the kinetic depinning field.  
This shows that, one can tune the values of static and kinetic depinning fields and make them equal by tuning the current density. Also it can be verified that the static and kinetic depinning currents are -58.8$\times$10$^{8}$ A/cm$^{2}$(label a) and -78.9$\times$10$^{8}$ A/cm$^2$(label b) respectively in the absence of field as shown in FIG.4.

\section{Conclusions}
In this paper, the pinning and depinning of a Neel-type transverse domain wall in the pinning field region or notch created by two symmetrical triangles are reported in the presence of field and current. The corresponding LLG equation with the adiabatic and non-adiabatic spin-transfer torques along with the effective field is solved analytically to derive the velocity, width and excitation angle of the domain wall. The displacement of the domain wall is obtained by numerically integrating the velocity equation using Sympson's 3/8 rule. The static and kinetic depinning fields in the absence of current have been found as 100 Oe and 36.7 Oe respectively for a pinning field strength 100 Oe in the range $0<x\leq23.52$ nm . Also, the static and kinetic depinning current densities have been found as -58.8$\times$10$^{8}$ A/cm$^{2}$ and -78.9$\times$10$^{8}$ A/cm$^{2}$ respectively for the same pinning field strength and range in the absence of the field. For the field driven case, the static depinning field is greater than the kinetic depinning field whereas for the current driven case, the static depinning current density is smaller than the kinetic depinning current density. The velocity of the domain wall  to cross the notch is less in the current driven case(553.0 m/s) when compared to the field driven case(700.8 m/s).   At a particular value of the saturated current density(-54.5$\times$10$^{8}$ A/cm$^{2}$) both the depinning fields are equal and below -54.5$\times$10$^{8}$ A/cm$^{2}$, the static depinning field is greater than kinetic depinning field whereas above -54.5$\times$10$^{8}$ A/cm$^{2}$, the static depinning field is smaller than the kinetic depinning field. This work helps to improve the mechanism to control the motion of the domain wall by geometrical notches in ferromagnetic nanostrips.

\section{Acknowledgement}
The work of M.D and R.A forms part of a major DST project. P.S wishes to thank DST-SERB Fast track scheme for providing financial support.

\bibliographystyle{pramana}
\bibliography{references}
\end{document}